\documentstyle[epsfig]{aipproc}

\def\kms{{km s$^{-1}$}}

\def\ee #1 {\times 10^{#1}}
\def\ut #1 #2 {\hbox{\thinspace #1}^{#2}}
\def\u #1 {\hbox{\thinspace #1}}
\def\msol{\hbox{$\hbox{M}_\odot$}}

\def\figpage #1 #2 {\vfill\eject\pageno=#1\vglue 8.0in\line{\hfill #2}}

\begin{document}
\title{\vspace{-36pt} The Origin of the High-Energy  \\ 
Activity at the Galactic Center}

\author{\vspace{-8pt} Farhad Yusef-Zadeh$^*$, William Purcell$^*$ 
        \& Eric Gotthelf$^{\dagger}$}
\address{\vspace{-6pt} $^*$Dept. Physics and Astronomy, 
Northwestern University, Evanston, Ill. 60208\\
$^{\dagger}$NASA/GSFC, Greenbelt, MD 20771}

\maketitle

\vspace{-28pt}

\begin{abstract}
Recent X-ray and gamma-ray observations of the Galactic center region by
the ASCA and CGRO/EGRET instruments show evidence of 2--10 keV and $>1$ GeV
continuum emission as well as 6.7 and 6.4 keV line emission from the
inner $0.2^\circ$ of the Galactic center.  This region is also known to
host a bright nonthermal radio continuum source Sgr A East and a dense
molecular cloud M--0.02--0.07 known as the 50 km s$^{-1}$ cloud.  The 
oval-shaped 
nonthermal Sgr A East is physically interacting with M--0.02--0.07 at the
Galactic center. 
A comparison between the distribution of 
ionized,  synchrotron and neutral gas  
suggests a self-consistent interpretation of the high-energy activity at 
the Galactic center. Our preliminary analysis of the 
data suggest a 
  shock model of cosmic ray acceleration at
the site of the  interaction to explain the enhanced GeV 
$\gamma$-ray emission. We also 
address a number of issues related to the spatial correlation of the diffuse
radio and X-ray emitting gas as well as to the origin of the fluorescent 6.4
and 6.7 keV emission at the Galactic center.
\end{abstract}

\vspace{-8pt}

\section*{Introduction}

\vspace{-2pt}

\underline{\it Radio View of the Galactic Center}
Radio continuum observations of the inner 15$'$ of the 
Galactic center show two prominent 
radio continuum structures known as the Sgr A complex and the 
filamentary continuum  Arc.  
The Sgr A Complex consists of Sgr
 A East and  its halo as well as Sgr A$^*$ 
and its thermal orbiting gas Sgr A West.
 Sgr A$^*$ is unique and  considered by many to be 
a massive black hole with a mass of 10$^6 \msol $ at the Galactic 
center.   The oval-shaped structure  known as Sgr A East
 is thought to be the remnant of an explosion
located just behind the
Galactic center (Yusef-Zadeh and Morris 1987; Pedlar et al. 1989),
A number of authors (Khokhlov \& Melia 1996;   Mezger et
al. 1989; Yusef-Zadeh and Morris 1987)
 question the interpretation of Sgr A East as a standard
SNR.  Khokhlov \& Melia (1996) have considered that Sgr A East is the 
remnant of star that is tidally disrupted by a massive black hole, 
presumably by Sgr A$^*$. The explosion energy is estimated  to be 
an order of magnitude more than the energy released by a typical 
supernova. There is also considerable  evidence that this explosion 
occurred inside the dense molecular cloud M--0.02--0.07, thus 
depositing more than 4$\times10^{52}$ ergs in the ISM (Mezger et al. 
1989).  Recent discovery of OH(1720MHz) masers at the interface of the 
50 \kms\ molecular cloud and Sgr A East showed conclusively that these 
two are physically interacting with each other (Yusef-Zadeh et al. 
1996). 

On the largest scale, there is 
a diffuse 7--10$'$ halo of nonthermal continuum 
emission surrounding the oval-shaped  radio structure Sgr A 
East. The spectrum of the halo tends to be steeper than Sgr A East  and is 
primarily nonthermal with the energy 
spectral index of $>$3. The optical depth toward Sgr A East  
and the halo at low frequencies lead Pedlar et al. (1989) 
to consider  a mixture of both thermal
and nonthermal gas but  displaced to the front side of Sgr A East. 

The Arc is a nonthermal filamentary source 
located near l$\approx0.18^0$
and runs in the direction perpendicular to the Galactic plane. The 
filaments are linearly polarized showing evidence that they 
are tracing  magnetic field lines and emitting synchrotron radiation. 
The spectrum of the filaments is unusual in that it is flatter than typical 
nonthermal features. The energy spectrum of relativistic particles
have a spectral index of $\approx$1.6 in the radio wavelengths.
A number of Galactic center molecular clouds appear to outline the linear
filaments, prompting the hypothesis that the filaments and clouds are
physically interacting with each other, in which case the field strength is
estimated to be at least 1 mG in order for the filaments to resist
deflection at points of interaction with the clouds (Yusef-Zadeh and Morris
1987; Serabyn and Morris 1994). 


\vspace{6pt}

\underline{\it X-Ray View of the Galactic Center}
Recent ASCA observations of the Galactic center showed conclusively the 
evidence for diffuse X-ray emission arising from the inner 15$'$ of the 
Galactic center (Koyama et al. 1996). 
The continuum radiation is accounted for by thermal plasma having 
temperature of 10 keV. 
 The strongest 
continuum radiation from the 
inner 30$'$ in the energy band between 0.7 and 10 keV 
arises from  within the shell of Sgr A 
East and is somewhat elongated along the Galactic plane. 
The electron density 
and the thermal energy 
of thermal gas within the shell of Sgr A East are 
estimated to be 6 cm$^{-3}$ and 3$\times 10^{50}$ ergs, respectively (Koyama 
et al. 1996).  
Weak and diffuse emission  is 
also seen corresponding to the radio halo as well along the Galactic 
plane  both in the positive and negative longitudes,  but with low surface 
brightness. This weak emission extends over as far as 80 pc (33$'$) on either 
side of the Galactic center. The strongest diffuse emission beyond the Sgr A 
complex arises from the positive longitude side outlined by the nonthermal 
radio  filaments of the Arc. Figure 1 shows X-ray contours   
superimposed on  the radio image displaying the Sgr A 
East shell, its halo and the filaments in the Arc. 
The electron density and the thermal energy 
of the hot gas outside the Sgr A 
complex are  estimated to be about 0.3--0.4 cm$^{-3}$ and 
0.5--1$\times10^{53}$ ergs, respectively (Koyama et al. 1996).

One of the more fascinating aspect of ASCA observations of the Galactic 
center is the evidence of 6.4 keV  emission peaking on two molecular 
clouds in the region between the Arc and the Sgr A complex and in Sgr 
B2.  This fluorescent K$\alpha$ line emission results from the K-shell 
photoionization of iron atoms. 


\vspace{6pt}

\underline{\it Gamma-Ray View of the Galactic Center}
Recent report of high energy (30 MeV -- 30 GeV) 
continuum 
emission from the Galactic center based on EGRET observations (Mattox 
1997) indicate a source with a luminosity of 5$\times10^{36}$ erg 
s$^{-1}$ at the distance of the galactic center. The source is situated within 
0.2$^0$ of the Galactic center and could be either compact or diffuse 
within 100 pc of the Galactic center. The energy spectrum of this source 
is fit by a power law  having an index of 1.7 which is harder than 
typical EGRET sources in the Galactic plane. This source has been considered 
to be associated with the Arc (Pohl 1997),  or with Sgr A$^*$, or 
possibly 
to have a diffuse origin (Thompson et al. 1996; Mattox  1997).

\vspace{-8pt}

\section*{Discussion}

\vspace{-2pt}

\underline{\it Sgr A East as the Source of High-Energy Activity}
The  hypothesis that we are 
considering  involves Sgr A East, the most energetic source 
in the Galactic center region. This source has been considered 
to be due to an unusual explosion, perhaps a Seyfert-like 
 activity as seen in the nucleus of spiral galaxies 
(Pedlar et al. 1989).
A number of observations indicate that Sgr A East 
is  unusual and more energetic than a typical 
supernova, having released  more than $5\times10^{52}$ 
ergs into the Galactic center region (e.g. Mezger et al. 1989). 

Since the interaction of Sgr A East with the 50 \kms\ molecular cloud
is well established, 
the high-energy cosmic rays responsible for radio, X-ray and 
$\gamma$-ray emission could  be 
generated at the site of the interaction  of Sgr A East and its 
molecular cloud before diffusing  out  
along the Galactic plane.  
In this hypothesis, 
the EGRET source is considered to be diffuse and the $\gamma$-ray 
spectrum is due to accelerated cosmic rays at the site of the 
interaction of the explosive event with the giant molecular cloud.
The western edge of the 50 \kms\ 
giant molecular cloud is interacting with the Sgr A East shell
whereas the eastern edge of the cloud appears to be outlined by the 
nonthermal linear filaments that cross the Galactic plane near 
l$\approx0.2^0$. Figure 2
shows the overall distribution of  $^{13}$CO gas between 30 and 50 km s$^{-1}$
 (Bally et al. 1988) which appears to 
be correlated with the hot X-ray emitting gas displayed as contours in 
Figure 1.
In addition, the distribution of 6.4 keV emission from the inner 30$'$ of the 
Galactic center, as discussed below, 
appears to coincide with the peaks of two 
molecular gas clouds (Koyama et al. 1996) in M--0.02--0.07.
These morphological correlations strongly suggest that Sgr A East, 
the 50 \kms\ molecular cloud, the hot X-ray emitting gas and the 6.4 keV 
emission are physically associated implying that Sgr A East is 
responsible for the high-energy activity at the Galactic center. 

  There is considerable evidence 
that nonthermal particles at high energies produce synchrotron emission 
from Sgr A East and the Arc as well as the continuum 
EGRET $\gamma$-ray source (2EGJ1746-2852) seen at the 
Galactic center. There is also conclusive 
evidence that shocks exit 
at the interface of the 50 \kms\ molecular cloud and Sgr A East. Thus, it 
is plausible that the 
low-energy cosmic ray particles must exist 
from extrapolation of high-energy nonthermal particles.
In fact some of the brightest supernova remnants interacting with 
molecular clouds and showing OH maser shocks at 1720 MHz 
appear to have  EGRET counterparts (Esposito et al. 1996).

The eastern edge of the Sgr A East lies a high density molecular core 
with a density of 1--2$\times10^6$ cm$^{-3}$
and a string of HII regions excited by massive stars (e.g. Serabyn et al. 
1992). Both these features are associated with sites of star 
formation in the 50 \kms\ cloud. The cluster of hot stars and the 
cavity of ionized gas surrounded by the circumnuclear ring at 
the Galactic center begs the question of the association of these
features with another site of star formation within the 50 \kms\ cloud. 
We sketch a scenario in which 
the cluster of hot stars IRS 16 and the circumnuclear ring of molecular gas 
are remnants of an episode of star formation that took place 
in the 50 \kms\ cloud. In 
this picture, the stars and the gas are captured at the Galactic center 
as the giant molecular cloud passed the Galactic center and is 
presently behind Sgr A West.


\vspace{6pt}

\underline{\it The Nature of 10$^7$--10$^8$ K Emission}
The 50 \kms\ molecular is  distributed along the positive longitude 
side of the plane but immediately 
behind the Galactic center.
The front side 
of the cloud is assumed to face Sgr A$^*$ and the 
shell of Sgr A East outlines the compressed molecular gas as a result 
of the shock wave driving into the the dense medium of M--0.02--0.07.
The halo of Sgr A East  seen both in the radio and X-ray wavelengths 
is a secondary manifestation of the explosion.
 The hot gas and 
the accelerated cosmic rays that are produced at the site of the shock 
leak preferentially toward the Galactic center because 
Sgr A East is bounded  by the giant molecular cloud. The X-ray emitting 
gas is most concentrated within the shell of Sgr A East resembling the 
structure of composite supernova remnants. A strong morphological 
correlation between the nonthermal radio features, the Arc and Sgr A 
East and its halo,  and the hot and cold thermal emitting gas as noted
in contours of Figures 1 and 2 indicate strongly that the X-ray gas has 
a nonthermal component and  is mixed with thermal gas at a temperature of 10 
keV.    In fact, there is evidence of a hard tail in the diffuse X-ray spectrum 
of the inner region of the Galaxy
 based on Ginga observations (Yamasaki et al. 
1996).

 The hot plasma at this temperature can not be confined by the
gravitational potential in the Galactic center, thus an alternative mechanism 
is required to constrain the hot gas in the region.  The morphology
of X-ray and radio emitting gas suggests that the X-ray gas follows the
magnetized filaments before reaching the Arc. Under the assumption that the
magnetic pressure of the Arc is strong enough to confine much of the
hot plasma, the estimated field strength has to be greater than 0.5 mG
for an electron density of 0.3 cm$^{-3}$ and a temperature of 10 keV.


\vspace{6pt}

\underline{\it The Nature of the 6.4 keV Emission}
Koyama et al. (1996) argue that the 
hot plasma can not account for exciting the  6.4 keV line emission since 
the observed X-ray luminosity is an order of magnitude less than 
the observed 6.4 keV line emission. Recent theoretical work by Borkowski 
and Szymkowiak (1997), however show that thermal electrons with few to 10 keV 
energies can penetrate interstellar dust grains and produce 
fluorescent K$\alpha$ emission  through the K shell ionization. The 
advantage of this mechanism over the excitation of K$\alpha$ of Fe 
in the gas phase is that the dust grains are abundant in the 
content of their heavy atoms. These authors consider a non-equilibrium 
ionization model to account for X-ray spectra of hot plasma mixed in with 
neutral atoms of heavy elements. 
The ionization parameter $\tau$ is the  product of electron density 
n$_e$ and time t. In the case when $\tau$ is 
less than 10$^{13}$ cm$^{-3}$ s all ions are 
in equilibrium and the contribution 
of Fe K$\alpha$ emission from dust grains is minimal. However, for the 
case when the plasma is underionized before the final equilibrium state, 
the Fe K$\alpha$ emission could be  totally due to dust grains. 
Since the observed 6.4 keV and 
6.7 keV line intensities arise from the  dense molecular clouds outside
the Sgr A East complex, 
it is plausible 
that the thermal electrons of the plasma with a 10 keV temperature is
exciting the  Fe K$\alpha$ of dust grains in selected clouds where the
non-equilibrium ionization model is applicable.

We believe that the two prominent clouds toward Sgr A East and Sgr A
West (the circumnuclear disk) at the Galactic center, namely the 50
\kms\ and the circumnuclear neutral gas orbiting Sgr A$^*$, do not show
any 6.4 keV line emission because the ionization parameter is
close to 10$^{13}$ cm$^{-3}$ s.  With the expansion time scale of
5$\times10^4$ yrs over 80 pcs and the electron density of 6 cm$^{-3}$
in Sgr A East, $\tau$=10$^{-13}$ cm$^{-3}$. However, the region where
the Fe K$\alpha$ is seen, the electron density of the hot plasma is
much lower than 6 cm$^{-3}$ and therefore the non-equilibrium
ionization model is applicable.

Ginga results have shown  a hard tail in the spectrum of the inner 
degree of the Galactic center (Yamauchi et al. 1990).
The  hard tail beyond 10 keV can be important 
in explaining  the nature of 6.4 keV emission 
and the strong correlation seen between CO and 2-10 keV 
spatial distributions and  in determining 
whether the low-energy cosmic ray electrons are responsible for heating 
of molecular clouds near the Galactic center.

\underline{\it Summary}
A  number of different observations indicate strongly the
co-existence of nonthermal and thermal gas in the inner 50 pc
of the Galaxy. This mixture of gas 
can be accounted for in a self-consistent fashion 
by assuming that the center of the high-energy activity is due to the unusual 
explosion of Sgr A East which is expanding inside 
the 50 \kms  giant molecular cloud.
In this picture, the high-energy ($\approx$GeV) 
cosmic rays accelerated by the explosion are responsible for the 
$\gamma$-ray emission as detected by CGRO/EGRET, the 
nonthermal radio continuum emission from Sgr A East and its halo, and 
for illumination of the magnetic field lines of the Arc. On the other 
hand, the low-energy 
($<$ 1 MeV) cosmic-ray particles are considered 
to heat the giant 50 \kms molecular cloud and produce the observed
diffuse X-ray emission, the 6.4 keV and the hard tail observed in 
the Ginga spectrum. The strong X-ray emission arising  simultaneously 
from the Sgr A East shell  and from the interior of  the shell 
shows a resemblance to the centrally peaked morphology of 
composite SNR's such as W44 (e.g. Rho et al. 1994). 
A more detailed account of 
this model will be given elsewhere.

\begin{figure} 
\vspace{10pt}
\caption{Radiograph of the $\lambda$20cm continuum image superimposed on
X-ray contours in the energy range between 0.7 and 10 keV 
with a resolution of 1'. The ASCA data was kindly provided by 
Dr. Koyama and Dr. Maeda.}
\label{fig1}
\end{figure}

\begin{figure} 
\vspace{10pt}
\caption{Using the 1950 coordinates, this figure is simialr 
to Figure 1 except that the contours 
represent the distribution of $^{13}$CO emission in the velocity range 
between 30 and 50 km s$^{-1}$ having similar resolution to that 
of X-ray data shown in Figure 1 (Yusef-Zadeh 1986; Bally et al. 1988). }
\label{fig2}
\end{figure}

\end{document}